\documentclass[12pt,a4paper]{article}
\usepackage{amsmath,amscd}
\usepackage{amssymb}
\usepackage{bm}
\usepackage{cite}  
\usepackage{comment}
\usepackage{graphicx}
\usepackage{hyperref}
\usepackage{makeidx}
\usepackage{mathpazo} 
\usepackage{mathtools}
\usepackage{multimedia}
\usepackage{slashed}
\usepackage[small,nohug,heads=vee]{diagrams}
\usepackage[usenames,dvipsnames]{xcolor}
\usepackage{xfrac}

\def\func#1{\mathop{\rm #1}\nolimits}
\definecolor{light-gray}{gray}{0.80}

\diagramstyle[labelstyle=\scriptstyle]

\newcommand{\cm}{\mathcal{M}}
\newcommand\eq{=&&\hspace{-18pt}}
\newcommand\p{\hspace{1pt}}
\newcommand\strt[1]{\rule[-#1pt]{0pt}{#1pt}}

\newcommand\w{\hspace{4pt}}

\usepackage{fancyhdr}
 \renewcommand{\maketitle}{
     \begin{center}
       \Large
         {\bf An Evolving Spacetime Metric Induced \\ by a `Static'
	 Source\footnote{\noindent Appears in \textit{Symmetry} \textbf{2023},
	 15, 1381. \url{https://doi.org/10.3390/sym15071381}}}
         \vskip .3 true cm
       \small
         Martin Land \\
         \vskip .3 true cm
         Department of Computer Science \\
         Hadassah College \\
         37 HaNevi'im Street, Jerusalem \\
 email: martin@hac.ac.il
       \end{center}
       \vskip .5 true cm
 }
\begin{document}
\title{}
\author{}
\maketitle


%
\begin{abstract}
In a series of recent papers we developed 
a formulation of general relativity in which spacetime and the dynamics of matter 
evolve with a Poincar\'e invariant parameter $\tau$.  
%
In this paper, we apply the formalism to derive the metric
induced by a `static' event evolving uniformly along its $t$-axis at the spatial
origin ${\mathbf x} = 0$.
The metric is shown to vary with $t$ and $\tau$, as well as spatial distance
$r$, taking its maximum value for a test particle at the retarded time $\tau = t
- r/c$.  
In the resulting picture, an event localized in space and time produces a metric
field similarly localized, where both evolve in $\tau$. 
We first derive this metric as a solution to the wave equation in
linearized field theory, and discuss its limitations by studying the
geodesic motion it produces for an evolving event.
By then examining this solution in the 4+1 formalism, which poses an initial value
problem for the metric under $\tau$-evolution, we clarify these limitations and
indicate how they may be overcome in a solution to the
full nonlinear field equations.

\vskip 12pt \noindent keywords: general relativity; the problem of time; Stueckelberg--Horwitz--Piron theory; parameterized relativistic
mechanics
\end{abstract}

\baselineskip7mm 
\parindent=0cm \parskip=10pt

\section{Introduction}
\label{intro}

The {4+1} 
 formalism~\cite{AS,sym12101721,Land_2021,universe8030185,Land_2023} in
general relativity (GR) poses an initial value for the spacetime metric in which
evolution of fields and matter is parameterized by a Poincar\'e invariant
chronological time $\tau$. 
Parameterization in proper time was introduced in 1937 by \hbox{Fock
~\cite{Fock}}
in their manifestly covariant electrodynamics. 
However, in~1941, Stueckelberg~\cite{Stueckelberg-1,Stueckelberg-2} showed that
neither coordinate time $t$ nor the proper time $ds = \sqrt{-dx^\mu dx_\mu}$
can be used as a chronological evolution parameter in an electrodynamics that
accounts for pair creation/annihilation processes.
Instead, to~describe antiparticles as particles whose trajectory reverses
direction in coordinate time $t$, he introduced a strictly monotonic evolution parameter $\tau$,
independent of phase space and external to the spacetime~manifold.  

Piron and Horwitz~\cite{HP} generalized Stueckelberg's formalism, constructing
a relativistic canonical many-body theory
~\cite{bound-1,bound-2,bound-3,bound-4,bound-5} 
with Lorentz scalar Hamiltonian.  
By including $\tau $ in the U(1) gauge freedom (but not the spacetime manifold),
the Stueckelberg--Horwitz--Piron (SHP) formalism
in flat spacetime~\cite{saad,rel-qm,manybody,RCM} provides an electrodynamic theory of events
interacting through five gauge potentials.
The evolution of a localized spacetime event 
induces a field
acting on a localized remote event, 
through an interaction
synchronized by the chronological \hbox{time $\tau$}, and~recovering Maxwell
electrodynamics in $\tau$-equilibrium.

The structure of these interactions suggests a higher symmetry such as {O(3,2)} 
 or
{O(4,1)} for free fields, but~the observed Lorentz invariance of spacetime
requires that any 5D symmetry break to 4D tensor and scalar representations of
{O(3,1)} in the presence of matter.   
A similar conflict of symmetries is familiar from classical acoustics, where the
pressure wave equation appears invariant under Lorentz-like transformations, but~no relativistic effects are expected for observers approaching the speed of
sound.  
These considerations are a guiding principle in extension of the formalism to
general~relativity.

Horwitz has extended the SHP framework to curved spacetime~\cite{SHPGR,SHPGR2},
developing a classical and quantum theory of interacting event evolution 
in a background metric $g_{\mu\nu}(x)$.
As a many-body theory with $\tau$-evolution, the~scalar event density $\rho(x,\tau) $ and
energy-momentum tensor $T_{\mu\nu}(x,\tau)$ naturally become explicitly $\tau$-dependent.
In keeping with Wheeler's characterization~\cite{wheeler_bio} of Einstein
gravitation as ``spacetime tells matter how to move; matter tells spacetime how to
curve'', the $\tau$-dependent matter distribution must be reflected in
a $\tau$-dependent local metric $\gamma_{\mu\nu}(x,\tau)$. 
Particle dynamics in such a metric spacetime may differ from standard GR---some
details are indicated in Section~\ref{event}.
%
%
As in flat space electrodynamics, the~free fields of GR---the geometrical
structures---enjoy 5D spacetime and gauge symmetries, but~the spacetime symmetry must break
to {O(3,1)} in the presence of matter.
Because the metric evolution is parameterized by the external parameter $\tau$
and the matter evolution is determined by an {O(3,1)} scalar Hamiltonian,
there is no conflict with the diffeomorphism invariance of
general relativity.   
Some details of the {4+1} method are reviewed in Sections~\ref{local} and
\ref{weak}.

Several simple examples of the {4+1} formalism were given in previous papers,
but these did not involve a source event evolving along a localized trajectory.
In this paper, we study the field induced by a localized event, with~the goal
of describing a $\tau$-localized metric and the gravitational field it produces
on a remote localized event. 
We proceed in analogy to SHP electrodynamics where a particle is modeled as an
ensemble of events~\cite{RCM} located at {${\mathbf x} = 0$} 
 in space, but~narrowly distributed along the $t$-axis.
The 5D wave equation leads to the Coulomb potential~\cite{RCM} in the form   
\begin{equation}
a_0\left( x,\tau \right) = - \frac{\varphi \left( t - r/c - \tau \right)}{r} +
o\left( \frac{1}{r^2} \right) ,
\label{coulomb}
\end{equation}
where $\varphi (s)$ is the distribution on the $t$-axis, with~maximum at $\varphi
(0)$ and normalized as $ \int d\tau \p \varphi (\tau) = 1$.  
At long distances, the~higher order term may be neglected.
A test event at spatial distance $r$ will thus experience a potential localized
around $\tau = t_{R} = t - r/c $, the~retarded time at which the source event produced
the field, where $c$ is the speed of light.
%
The general Li\'enard--Wiechert potentials induced by an event on an arbitrary
trajectory appear in their usual form~\cite{RCM}, but~multiplied by $\varphi\left( t - r/c - \tau \right)$.  

For the gravitational field, we similarly consider the metric induced by a
`static' event evolving uniformly along the $t$-axis in its rest frame, fixed at
the spatial origin ${\mathbf x} = 0$,
leading to an event current and mass-energy-momentum tensor
$T^{\mu\nu}(x,\tau)$. 
In Section~\ref{5D-wave}, we use this tensor as the source of a wave equation 
in linearized GR, and~derive a metric that varies with $t$ and
$\tau$, as~well as spatial distance $r$. 
Neglecting the higher order contribution as in electrodynamics, a~test particle
with coordinates \hbox{$x = \left( x^0(\tau), {\mathbf x}(\tau) \right)$}
experiences a metric that takes its maximum at $\tau = t - \vert {\mathbf x}
\vert /c$.  
However, unlike the flat space motion of an event under the electrodynamic Lorentz force,
the geodesic equations for an event moving in this metric differ from our
expectations, suggesting that localization along the $t$-axis
may cause the gravitational force to change sign.
We show that this issue follows from the structure of the Green's function for
the wave equation in linearized GR and will obtain for any $t$-dependent event
density.

In Section~\ref{evolution}, we examine this solution in the 4+1 formalism, which
poses an exact initial value problem for the metric under $\tau$-evolution.  
In this context, neglecting the higher-order term is seen to contradict the
assumption of an evolving metric, clarifying the limitations of the linearized
method.  
We pose the problem of an evolving metric produced by an evolving source
narrowly distributed in spacetime in terms of the full nonlinear field equations
and discuss the additional complexities associated with this system.   
Finally, Section~\ref{conclusion} is devoted to conclusions and discussion.
In a subsequent paper, numerical solutions to the initial value problem will be
discussed.

\section{Review of General Relativity with Invariant~Evolution}
\label{SHP}
\unskip

\subsection{Gauge and Spacetime~Symmetries}
\label{em}

In a flat Minkowski spacetime with $\eta_{\mu\nu} = \text{diag}\left( -1,1,1,1\right) $,
the free particle action
\begin{equation}
S = \int d\tau ~\frac{1}{2}M\dot{x}^{\mu }\dot{x}_{\mu },
\end{equation}
is made maximally U(1) gauge invariant~\cite{saad} by introducing five gauge
fields as
\begin{eqnarray}
S_{\text{SHP}}
\eq \int d\tau ~\frac{1}{2}M\dot{x}^{\mu }\dot{x}_{\mu }+\frac{e
}{c}\dot{x}^{\mu }a_{\mu }\big( x,\tau \big) +\frac{e}{c}
c_{5}a_{5}\big( x,\tau \big),
\label{SM} \\
\eq \int d\tau ~\frac{1}{2}M\dot{x}^{\mu }\dot{x}_{\mu }
+\frac{e }{c}\dot{x}^{\beta }a_{\beta }\big( x,\tau \big),
\label{SSHP}
\end{eqnarray}
where we introduce $x^5 = c_5 \tau$ in analogy to $x^0 = c t$ and
partition Greek indices such that
\begin{equation}
\alpha,\beta,\gamma,\delta = 0,1,2,3,5
\qquad \qquad
\lambda, \mu,\nu, \rho \ldots = 0,1,2,3 . 
\end{equation}

{This} 
 action enjoys the 5D gauge invariance $ a_{\alpha }\left(x ,\tau \right)
\longrightarrow  a_{\alpha }\left(x ,\tau \right) + \partial_\alpha \Lambda
(x,\tau)$, but~because $\dot{x}^{\mu }\dot{x}_{\mu }$, $\dot{x}^{\mu }a_{\mu }$,
and $a_{5}$ are O(3,1) scalars, its spacetime symmetry is restricted to 4D.   
As a guide to posing field equations for a $\tau$-dependent metric in curved
spacetime, we may consider Equation~(\ref{SSHP}) as a standard 5D action in which we
break the symmetry of the matter term by imposing the constraint $\dot x^5
\equiv c_5$ and making the replacement $\dot{x}^\alpha
\dot{x}_\alpha \longrightarrow \dot{x}^\mu \dot{x}_{\mu }$ in the kinetic term,
restricting the phase space to $ \left( x^\mu, \dot x^\mu \right) $. 
The electrodynamics associated with the symmetry-broken action differ in
significant ways from standard Maxwell theory in 5D.
In particular, the~Lorentz force~\cite{RCM}
\begin{equation}
M\ddot{x}_{\mu } = \frac{e}{c}\dot{x}^{\beta }f_{\mu \beta } 
%
%
\qquad \qquad 
\frac{d}{d\tau }\left( - \frac{1}{2}M\dot{x}^{\mu }\dot{x}_{\mu }\right)
= c_{5}\frac{e}{c}\dot{x}^{\mu }f_{5 \mu }
\qquad \qquad f_{\alpha\beta} = \partial_\alpha a_\beta - \partial_\beta
a_\alpha,
\label{L-2}
\end{equation}
permits mass exchange between particles and fields, while leaving the total mass,
energy, and~momentum of particles and fields conserved. 
Compatibility with standard electrodynamic phenomenology places restrictions on $c_5 / c
\ll 1$ but the strict limit $c_5 \longrightarrow 0$ produces a $\tau$-equilibrium
~\cite{RCM} that recovers standard Maxwell~theory.

Field dynamics are determined by a kinetic term of the type
\begin{equation}
S_{\text{field}} = \int d\tau \p d^4x f^{\alpha\beta} (x,\tau) f_{\alpha\beta}
(x,\tau) ,
\label{f-kin}
\end{equation}
where $f_{\mu\nu}$ is a second rank tensor, while $f_{5\mu}$ is a vector field
strength, because~the 5-index signifies an {O(3,1)} scalar quantity.
Raising the 5-index in (\ref{f-kin}) suggests a 5D flat space metric 
\begin{equation}
\eta_{\alpha\beta} = \text{diag} \left( -1,1,1,1,\sigma \right)  ,
\label{flat}
\end{equation}
where $\sigma = \pm 1$.
But expanding
\begin{equation}
f^{\alpha\beta} (x,\tau) f_{\alpha\beta} (x,\tau) =
f^{\mu\nu} (x,\tau) f_{\mu\nu} (x,\tau) + 2 \sigma f^\mu_{\ 5} (x,\tau) f_{\mu
5} (x,\tau) ,
\end{equation}
we may regard $\sigma$ as the choice of sign for the vector-vector interaction,
with no inherent significance for the geometry of spacetime.

Following these considerations, we approach the construction of a
$\tau$-dependent GR by embedding 4D spacetime $\cm$ in a 5D pseudo-spacetime
$\cm_5 = \cm \times R$ with coordinates $X^\alpha = \left( x^\mu, c_5 \tau
\right)$ and a metric $g_{\alpha\beta}(x,\tau)$ determined by the standard 5D
Einstein field equations on $\cm_5$. 
By performing the embedding in a vielbein frame~\cite{Yepez}, we may specify the
metric as (\ref{flat}) and break the 5D spacetime symmetry to {O(3,1)} at the
source, by~correcting the flat space metric for the quintrad for the matter
terms under the replacement 
\begin{equation}
\eta_{ab} = \text{diag} \left( -1,1,1,1,\sigma \right) \
\longrightarrow \ \widehat \eta_{ab} = \text{diag} \left( -1,1,1,1,0 \right) ,
\label{replace}
\end{equation}
as in electrodynamics.
The symmetry-broken field equations are transformed into the coordinate frame as
\begin{equation}
R_{\alpha \beta }=\frac{8\pi G}{c^{4}}\left( T_{\alpha \beta }-\frac{1}{2}
\widehat g_{\alpha \beta }\widehat{T}\right) ,
\label{field-1}
\end{equation}
using the known vielbein field. 
The LHS of Equation~(\ref{field-1}) enjoys 5D gauge and spacetime
symmetries, while the RHS is O(3,1) covariant.
Generalizing the {3+1} formalism in geometrodynamics
~\cite{Gourgoulhon,Bertschinger,Blau,ADM} to {4+1}, we take advantage of the
natural foliation of $\cm_5$ into 4D equal-$\tau$ spacetimes homeomorphic to $\cm$.
Standard techniques in the theory of embedded surfaces enable us to
extract an initial value problem in the 4D spacetime sector, describing the
$\tau$-evolution of metric $\gamma_{\mu\nu}(x,\tau)$ and the extrinsic curvature
$K_{\mu\nu}(x,\tau)$ that accounts for aspects of the 5D connection
$\Gamma^\gamma_{\w\alpha \beta }$ not contained in the 4D~metric.

\subsection{Event Dynamics in Curved~Spacetime}
\label{event}

Applying the Euler--Lagrange equations to the Lagrangian
\begin{equation}
L=\frac{1}{2}mg_{\alpha \beta }\dot{x}^{\alpha }\dot{x}^{\beta } ,
\end{equation}
we obtain the 5D geodesic equations for an event $x^\gamma(\tau) $
\begin{equation}
\frac{D\dot{x}^{\gamma }}{D\tau } = \ddot{x}^{\gamma }+\Gamma _{\alpha \beta
}^{\gamma }\dot{x}^{\alpha }\dot{x}^{\beta } ,
%
\end{equation}
with Christoffel symbols 
\begin{equation}
\Gamma _{\beta \gamma }^{\alpha }=\frac{1}{2}g^{\alpha \delta }\left( \frac{
\partial g_{\delta \beta }}{\partial x^{\gamma }}+\frac{
\partial g_{\delta \gamma }}{\partial x^{\beta }}-\frac{
\partial g_{\beta \gamma }}{\partial x^{\delta }}\right) . 
\end{equation}

We break the 5D symmetry to {O(3,1)} by asserting
\begin{equation}
x^{5}=c_{5}\tau \ \ \longrightarrow \ \ \dot{x}^{5}=c_{5} \ \ \longrightarrow \
\ \ddot{x}^{5}=0 ,
\end{equation}
as an a priori constraint.
The dynamical system is now described by the equations  
\begin{eqnarray}
\frac{D\dot{x}^{\mu }}{D\tau } \eq  \ddot{x}^{\mu }+\Gamma _{\alpha \beta
}^{\mu }\dot{x}^{\alpha }\dot{x}^{\beta }
= \ddot{x}^{\mu }+\Gamma _{\nu \sigma }^{\mu }\dot{x}^{\nu }\dot{x}^{\sigma
}+2c_{5}\Gamma _{5\nu }^{\mu }\dot{x}^{\nu }+c_{5}^{2}\Gamma _{55}^{\mu } = 0 ,
\strt{12}
\\
\frac{D\dot{x}^{5}}{D\tau }\eq \ddot{x}^{5}\equiv 0 ,
\end{eqnarray}
which under an appropriate metric field is consistent with the symmetries of
matter, and~recovers standard GR when $g_{5\alpha} = 0$ and $\partial_\tau
g_{\mu\nu} = 0$. 
Defining the canonical momentum 
\begin{equation}
p_{\mu }  = \frac{\partial L}{\partial \dot{x}^{\mu }}
=m\left( g_{\mu \nu }\dot{x}^{\nu }+c_{5}g_{\mu 5}\right)  ,
\end{equation}
the Hamiltonian is
\begin{equation}
K  = \frac{1}{2M}p^{2}+\frac{1}{2}c_{5}g_{55}g^{5\mu }p_{\mu }-\frac{1}{2}%
c_{5}g_{5\mu }g^{\mu \lambda }p_{\lambda }+\frac{1}{2}Mc_{5}^{2}g_{5\mu
}g^{\mu \lambda }g_{\lambda 5}+\frac{1}{2}Mc_{5}^{2}g_{55} ,
\end{equation}
which takes the recognizable form
\begin{equation}
K=\frac{1}{2m}p^{\mu}p_{\mu}+\frac{1}{2}mc_{5}^{2} \ g_{55} ,
\end{equation}
if $g^{5\mu }=0$, with~$g_{55} (x,\tau)$ playing the role of a
$\tau$-dependent potential on 4D spacetime.
The canonical equations of motion are
\begin{equation}
\dot{x}^{\mu }=\frac{dx^{\mu }}{d\tau }=\frac{\partial K}{\partial p_{\mu }}
\qquad \qquad \dot{p}_{\mu }=\frac{dp_{\mu }}{d\tau }=-\frac{\partial K}{
\partial x^{\mu }} ,
\end{equation}
and since $p_{5}\equiv 0$, the~Poisson bracket is 
\begin{equation}
\left\{ F,G\right\} =\frac{\partial F}{\partial x^{\alpha }}\frac{\partial G
}{\partial p_{\alpha }}-\frac{\partial F}{\partial p_{\alpha }}\frac{
\partial G}{\partial x^{\alpha }}=\frac{\partial F}{\partial x^{\mu }}\frac{
\partial G}{\partial p_{\mu }}-\frac{\partial F}{\partial p_{\mu }}\frac{
\partial G}{\partial x^{\mu }} ,
\end{equation}
so for any scalar function $F\left( x,p,\tau \right) $ on phase
space
\begin{equation}
\frac{dF}{d\tau } = \left\{ F,K\right\} +\frac{\partial F}{\partial \tau } ,
\end{equation}
generalizing the nonrelativistic result.
Therefore, the~Hamiltonian is conserved unless $K$ depends
explicitly on $\tau $ through $g_{\alpha \beta }\left( x,\tau \right) $.  
We note that even when $K$ is a constant of the motion, the~4D mass
$p^{\mu}p_{\mu} / 2m$ may vary under $g_{55}$.

As we showed in~\cite{sym12101721}, mass variation can appear in the Newtonian
approximation through $\tau$-dependence of the metric. 
Expanding the geodesic equations as
\begin{equation}
0 = \ddot{x}^{\mu }+\Gamma _{00}^{\mu }\dot{x}^{0}\dot{x}^{0}+2\Gamma
_{i0}^{\mu }\dot{x}^{i}\dot{x}^{0}+\Gamma _{ij}^{\mu }\dot{x}^{i}\dot{x}
^{j}+2c_{5}\Gamma _{50}^{\mu }\dot{x}^{0}+2c_{5}\Gamma _{5i}^{\mu }\dot{x}
^{i}+c_{5}^{2}\Gamma _{55}^{\mu } ,
\label{geo-eqn}
\end{equation}
we take $\partial_0 g_{\alpha\beta} = 0$ and neglect terms containing $\dot
x^i / c \ll 1$ for $i=1,2,3$, so that inserting the nonzero Christoffel symbols 
\begin{equation}
\begin{array}{lcl}
\Gamma _{00}^{\mu }=-\dfrac{1}{2}\eta ^{\mu \nu }\partial _{\nu }g_{00} & \qquad
&
\Gamma _{ij}^{\mu }=\dfrac{1}{2}\eta ^{\mu k}\left( \dfrac{\partial g_{ki}}{
\partial x^{j}}+\dfrac{\partial g_{kj}}{\partial x^{i}}-\dfrac{\partial
g_{ij}}{\partial x^{\nu }}\right)  ,\strt{16}
\\
\Gamma _{50}^{\mu }=\dfrac{1}{2}\eta ^{\mu \nu }\dfrac{\partial g_{\nu 0}}{
\partial x^{5}} & \qquad &
\Gamma _{55}^{\mu }=-\dfrac{1}{2}\eta ^{\mu \nu }\dfrac{\partial g_{55}}{
\partial x^{\nu }} ,
\end{array}
\end{equation}
the equations of motion reduce to
\begin{equation}
\frac{d^{2}t}{d\tau ^{2}}  = \frac{dt}{d\tau } \p \partial _{\tau }g_{00} \qquad \qquad 
\mathbf{\ddot{x}}  = \frac{1}{2}c^{2}\left( \frac{dt}{d\tau }\right)
^{2}\nabla g_{00}+\frac{1}{2}c_{5}^{2}\nabla g_{55} . 
\label{post-N}
\end{equation}

{Compared} to Newtonian gravity, 
which must obtain exactly when $\partial _{\tau
}g_{\mu \nu }$ and $\nabla g_{55}$ vanish, these become
\begin{equation}
\frac{d^{2}t}{d\tau ^{2}} = \frac{dt}{d\tau } \p  \partial_\tau \left(
\frac{2GM}{rc^{2}}\right) 
\qquad \qquad 
\frac{d^{2}\mathbf{x}}{d\tau ^{2}} = -\left( \frac{dt}{d\tau }\right) ^{2}
\frac{GM}{r^{2}}\mathbf{\hat{r }} +\frac{1}{2}c_{5}^{2}\nabla g_{55}  ,
\end{equation}
where $M$ is a mass parameter associated with the source and $G$ is the
gravitational constant.
Writing a perturbed source mass parameter $M = M_0 + \delta M (\tau) $ for the source and again
neglecting $\dot r / c$ we may solve the $t$ equation as  
\begin{equation}
\frac{dt}{d\tau }=\exp \left( \frac{2G}{rc^{2}} \p \delta M\right) ,
%
\label{t-N}
\end{equation}
so that in spherical coordinates, the~radial equation takes the form
\begin{equation}
\ddot{r}-\dfrac{L^{2}}{m^{2}r^{3}}+\exp \left( \frac{4G}{rc^{2}} \p \delta M
\right) \frac{GM_0}{r^{2}} = 0 ,
\label{r-N}
\end{equation}
where we take $\nabla g_{55} = 0$ and introduce the conserved angular momentum
\begin{equation}
L=mr^{2}\dot{\phi} . 
\end{equation}

{As} 
 required, Equations~(\ref{t-N}) and (\ref{r-N}) recover Newtonian gravitation
in the absence of the $\tau$-dependent source mass $\delta M$.
The Hamiltonian in this coordinate system is
\begin{equation}
K =  \dfrac{1}{2}mg_{\alpha \beta }\dot{x}^{\alpha }\dot{x}^{\beta } 
= -\dfrac{1}{2}mc^{2}\left( 1-\frac{2GM_0}{rc^{2}}\right) \exp \left( \frac{
4G}{rc^{2}} \p \delta M\right) +\dfrac{1}{2}m\dot{r}^{2}+\dfrac{1}{2}\frac{L^{2}}{
mr^{2}} ,
\end{equation}
with time derivative 
\begin{equation}
\dfrac{d}{d\tau }K = \exp \left( \frac{
4G}{rc^{2}} \p \delta M\right) \left( -\frac{Gm}{
r}+\frac{4G^{2}mM_0}{r^{2}c^{2}}\right) \frac{d}{d\tau} \delta M ,
\end{equation}
and as expected, the~Hamiltonian for the motion of this test particle is not
conserved in the presence of a variable mass gravitational source.
This may be interpreted as a transfer of mass across spacetime mediated by the
metric.

For non-thermodynamic dust (a distribution of geodesically evolving events
without mutual interaction), we define $\rho(x,\tau )$ as the number of events per
spacetime volume, and~write a 5-component event current with mass parameter $M$
\begin{equation}
j^{\alpha }\left( x,\tau \right) = M \rho (x,\tau )\dot{x}^{\alpha }(\tau)   ,
\label{E-19}
\end{equation}
with continuity equation
\begin{equation}
\nabla _{\alpha }j^{\alpha }=\frac{\partial j^{\alpha }}{\partial x^{\alpha }
}+j^{\gamma }\Gamma _{\gamma \alpha }^{\alpha }=\frac{\partial \rho }{
\partial \tau }+\nabla _{\mu }j^{\mu }=0   ,
\label{E-20}
\end{equation}
where the second equality holds because $j^5 = M c_5 \rho\left( x,\tau \right) $ is an
O(3,1) scalar and not the \hbox{5-component} of a vector with 5D symmetry. 
Generalizing the 4D energy-momentum tensor to
5D, the~mass-energy-momentum tensor
\begin{equation}
T^{\alpha \beta }=M \rho \dot{x}^{\alpha } \dot{x}^{\beta }\longrightarrow \left\{ 
\begin{array}{l}
T^{\mu \nu }=M \rho \dot{x}^{\mu }\dot{x}^{\nu } , 
\rule[-12pt]{0pt}{12pt} \\ 
T^{5\beta }=\dot{x}^{5}\dot{x}^{\beta }M \rho =c_{5}j^{\beta } , 
\end{array}
\right.   
\label{F-MEM}
\end{equation}
%
is conserved as
\begin{equation}
\nabla _{\alpha } T^{\alpha \beta } = 0 ,
\end{equation}
by virtue of the continuity and geodesic equations. 
The mass-energy-momentum tensor is thus a suitable {O(3,1)} covariant source for a 5D field
equation.

\subsection{Evolution of the Local~Metric}
\label{local}

As indicated in Section~\ref{em}, the derivation of field equations for
$g_{\mu\nu}(x,\tau)$ possessing the desired 5D gauge symmetries and 4D spacetime
symmetries relies heavily on the theory of embedded surfaces
~\cite{Gourgoulhon,Bertschinger,Blau}, the~{3+1} ADM formalism~\cite{ADM}, and~the generalization of these techniques to {4+1}
~\cite{sym12101721,Land_2021,universe8030185,Land_2023}. 
Here we provide a brief overview and refer the reader to the references for
details.

We approach the construction of GR with a
$\tau$-dependent metric by embedding 4D spacetime $\cm$ in a 5D pseudo-spacetime
$\cm_5 = \cm \times R$ with coordinates $X^\alpha = \left( x^\mu ,c_5
\tau\right) $. 
Because the Bianchi identity 
\begin{equation}
\nabla_\alpha G^{\alpha \beta}
= \nabla_\alpha \left( R^{\alpha \beta} - \dfrac{1}{2} g^{\alpha \beta}R \right) =
0 \qquad \qquad 
\nabla_\alpha X^\beta = \partial_\alpha  X^\beta+ X^\gamma
\Gamma^\beta_{\gamma\alpha} ,
%
%
\end{equation}
is independent of dimension~\cite{Weinberg} and the mass-energy-momentum tensor
(\ref{F-MEM}) is conserved, we may combine the Einstein tensor $G^{\alpha
\beta}$ and $T_{\alpha \beta}$ to write 
%
%
the field equations
\begin{equation}
R_{\alpha \beta} - \frac{1}{2} g_{\alpha \beta}R = k_G T_{\alpha \beta}  ,
\label{F-Einstein-5}
\end{equation}
for the metric $g_{\alpha\beta}(x,\tau)$ on $\cm_5$.
%

To break the spacetime symmetry of the field equations to O(3,1), we first transform from a
coordinate frame tangent to the manifold {$\cm_5$ } 
\begin{equation}
\mathbf{g}_{\alpha }=\partial_\alpha 
\qquad \qquad 
\mathbf{g}^{\alpha } = \mathbf{d} X^{\alpha }
\qquad \qquad 
\mathbf{g}_{\alpha }\cdot \mathbf{g}_{\beta }=g_{\alpha \beta }
\qquad \qquad 
\mathbf{g}^{\alpha }\cdot \mathbf{g}^{\beta }=g^{\alpha \beta } ,
\end{equation}
to the constant quintrad frame
\begin{equation}
\mathbf{e}_{a}\cdot \mathbf{e}_{b}=\eta _{ab} \qquad \qquad \mathbf{e}^{a}\cdot
\mathbf{e}^{b}=\eta ^{ab} \qquad \qquad 
\partial _{a} \mathbf{e}_{b} = \partial _{a}\mathbf{e}^{b} =0 ,
\end{equation}
where by convention Latin letters 
indicate a reference to the quintrad.
To facilitate foliation of $\cm_5$ into spacetime hypersurfaces $\Sigma_\tau$ of
equal-$\tau$, 
we extend the partition of coordinate indices to the quintrad indices, leading
to the combined index convention  
\begin{equation}
\begin{array}{lll}
\alpha,\beta,\gamma,\delta = 0,1,2,3,5 & \qquad \qquad  & \lambda, \mu,\nu,\rho
\ldots = 0,1,2,3 ,
\strt{12}\\
a,b,c,d,= 0,1,2,3,5 & \qquad  \qquad & k,l,m,n, \ldots = 0,1,2,3  ,
\end{array}
\end{equation}
where the five indices with respect to the quintrad frame are denoted $\bar 5 $.
The transformation between frames is provided by the vielbein field 
\begin{equation}
\begin{array}{ccc}
\mathbf{g}_{\mu }=E_{\mu }^{\hspace{4pt}k}\mathbf{e}_{k}+E_{\mu }^{\hspace{
4pt}\bar{5}}\mathbf{e}_{5}
& \qquad & 
\mathbf{e}_{k}=e_{\hspace{4pt}k}^{\mu }\mathbf{g}_{\mu }+e_{\hspace{4pt}
k}^{5}\mathbf{g}_{5}  ,\strt{12}
\\ 
\mathbf{g}_{5}=E_{5}^{\hspace{4pt}k}\mathbf{e}_{k}+E_{5}^{\hspace{4pt}\bar{5}}
\mathbf{e}_{5}
& \qquad &
\mathbf{e}_{5}=e_{\hspace{4pt}\bar{5}}^{\mu }\mathbf{g}_{\mu }+e_{
\hspace{4pt}\bar{5}}^{5}\mathbf{g}_{5} ,
\end{array}
\label{viel-1}
\end{equation}
%
%
where the spacetime hypersurface (quatrad) is spanned by the $\mathbf{e}_{k}$
while $\mathbf{e}_{5} $ points in the direction of $\tau$-evolution orthogonal
to $\Sigma_\tau$. 
Introducing the ADM parameterization 
\begin{equation}
\mathbf{g}_{5}=N^{\mu }\mathbf{g}_{\mu }+Nn   ,
\label{g_5}
\end{equation}
where $N^\mu$ generalizes the shift \hbox{3-vector}, 
$N$ is the lapse function with respect to $\tau$, and~$n = \mathbf{e}_{5}$ is
the unit normal, the~vielbein field becomes
\begin{equation}
\begin{array}{l}
E_{\alpha }^{\hspace{4pt}a}  = \delta _{\alpha }^{\mu }\delta _{k}^{a}E_{\mu
}^{\hspace{4pt}k}+\delta _{\alpha }^{5}\left( E_{\mu }^{\hspace{4pt}k}N^{\mu
}\delta _{k}^{a}+N\delta _{5}^{a}\right) , \strt{14} \\
e_{\hspace{4pt}a}^{\alpha }  = \delta _{a}^{k}\delta _{\mu }^{\alpha }e_{
\hspace{4pt}k}^{\mu }-\delta _{a}^{5}\delta _{\mu }^{\alpha }\dfrac{1}{N}
N^{\mu }+\delta _{a}^{5}\delta _{5}^{\alpha }\dfrac{1}{N} ,
\end{array}
\label{viel_sum}
\end{equation}
leading to the coordinated metric 
\begin{equation}
g_{\alpha \beta }=\left[ 
\begin{array}{cc}
\gamma _{\mu \nu } & N_{\mu } \rule[-14pt]{0pt}{14pt} \\ 
N_{\mu } & \sigma N^{2}+\gamma _{\mu \nu }N^{\mu }N^{\nu }
\end{array}
\right]  ,
%
\label{ADM_metric}
\end{equation}
which generalizes the ADM decomposition.
Since $N$ and $N^\mu$ are arbitrary functions acting as Lagrange multipliers
whose choice is comparable to gauge freedom~\cite{ADM},
the dynamical content in the vielbein field is contained entirely in the
spacetime vierbein field $E_\mu^{\hspace{4pt}k}$.

In the quintrad frame, the~Einstein equations take the form
\begin{equation}
R_{ab}-\frac{1}{2}\eta _{ab}R= k_G T_{ab} , 
%
\label{vielEin}
\end{equation}
and the spacetime symmetry may be broken to O(3,1) by making the replacement
(\ref{replace}) as
\begin{equation}
\eta _{ab} \longrightarrow \widehat{\eta}_{ab} = \delta^k_a\delta^l_b \p \p
\eta_{kl} ,
\end{equation}
in the matter terms, leading us to 
\begin{equation}
R_{ab} = k_G \left(
T_{ab}-\frac{1}{2}\widehat{\eta}_{ab}\widehat{T}\right) , 
\label{SymBrkViel}
\end{equation}
where $\widehat{T} = \widehat \eta^{\p ab} T_{ab} = \eta^{kl} T_{kl}$.
Using (\ref{viel_sum}), this expression transforms back to the coordinate frame
to provide the O(3,1) symmetric field equations
\begin{equation}
R_{\alpha \beta }= k_G\left( T_{\alpha \beta }-\frac{1}{2}
P_{\alpha \beta }\widehat{T}\right) , 
\label{SHP_field}
\end{equation}
with the transformed metric
\begin{equation}
\widehat{g}_{\alpha \beta }  = E_{\alpha }^{\hspace{4pt}a}E_{\beta }^{\hspace{4pt
}b}\p \widehat{\eta}_{ab} 
= g_{\alpha \beta }-\sigma n_{\alpha }n_{\beta }
= P_{\alpha \beta }  ,
\end{equation}
that acts as a projection operator from $\cm_5$ onto the 4D
spacetime hypersurface $\Sigma_\tau$ and thus breaks any higher symmetry to O(3,1). 

Given the foliation of the pseudo-spacetime into equal-$\tau$ spacetimes, the~initial value problem is found using $P_{\alpha\beta}$ to project the
geometrical structures from $\cm_5$ onto $\Sigma_\tau$:
\begin{enumerate}
	\item \label{1} The covariant derivative $D_{\alpha }$ on $ \Sigma_\tau$ 
	is found using $P_{\alpha\beta}$ to project the
	covariant derivative $\nabla_{\alpha }$ on $\cm_5$,   
	\item \label{2} The extrinsic curvature $K_{\alpha\beta}$ is defined by projecting
	the covariant derivative of the unit normal $n_\alpha$,
	\item \label{3} The projected curvature $\bar{R}_{\gamma\alpha\beta}^{\delta}$ on
	$\Sigma_\tau$ is defined through the
	non-commutation of projected covariant derivatives $D_{\alpha }$ and $D_{\beta }$,
	\item \label{4} 
	The Gauss relation is found by 
	decomposing the 5D curvature $R_{\gamma\alpha\beta }^{\delta}$ in
	terms of $\bar{R}_{\gamma\alpha\beta }^{\delta}$ and $K_{\alpha\beta}$,
	\item \label{5} The mass-energy-momentum tensor is decomposed through
	the projections
$$\qquad  \kappa  = n_{\alpha }n_{\beta }T^{\alpha \beta } \qquad 
p_{\beta }  = -n_{\alpha ^{\prime }}P_{\beta \beta ^{\prime }}T^{\alpha
^{\prime }\beta ^{\prime }} \qquad 
S_{\alpha \beta }  = P_{\alpha \alpha^\prime }P_{\beta \beta^\prime}T^{\alpha
^{\prime }\beta ^{\prime }}\qquad S = P^{\alpha \beta } S_{\alpha \beta } , $$
	\item \label{6} Projecting the 5D curvature $R_{\gamma\alpha\beta }^{\delta}$ on
	the unit normal $n_\alpha$ leads to the Codazzi relation providing a
	relationship between $K_{\alpha\beta}$ and $p_\alpha$,
	\item \label{7} Lie derivatives of $P_{\alpha\beta}$ and
	$K_{\alpha\beta}$ along the direction of $\tau$ evolution, given by
	the unit normal $n_\alpha$ in the coordinate frame, are combined
	with these ingredients, along with the O(3,1) symmetric field
Equation~(\ref{SHP_field}) to obtain $\tau$-evolution equations for $\gamma_{\mu\nu}$ and
	$K_{\mu\nu}$ and a pair of constraints on the initial~conditions.  

\end{enumerate}

The evolution equations are
\begin{equation}
\dfrac{1}{c_{5}}\partial_\tau \gamma _{\mu \nu }  =  {\mathcal{L}}_{{\mathbf N}}\,\gamma _{\mu
\nu }-2NK_{\mu \nu } ,
\label{der-gamma}
\end{equation}
\vspace{-12pt}
\begin{eqnarray}
\dfrac{1}{c_{5}}\partial_\tau K_{\mu \nu } \eq -D_{\mu }D_{\nu }N
+ {\mathcal{L}}_{\mathbf{N}} K_{\mu \nu } ,
\notag \\ 
&& +N\left\{ -\sigma \bar{R} _{\mu \nu }+KK_{\mu \nu }-2K_{\mu }^{\lambda
}K_{\nu \lambda }+\sigma k_G\left( S_{\mu \nu }-\frac{1}{2}P_{\mu \nu }S\right) 
\right\} , 
\label{der-K2}
\end{eqnarray}
where ${\mathcal{L}}_{\mathbf{N}}$ is the Lie derivative along $N^\mu$.
Their solutions must satisfy the Hamiltonian constraint
\begin{equation}
\bar{R}-\sigma \left( K^{2}-K^{\mu \nu }K_{\mu \nu }\right)
=
-k_G \left( S+\sigma \kappa \right)  , 
\label{c1}
\end{equation}
and the momentum constraint 
\begin{equation}
D_{\mu }K_{\nu }^{\mu }-D_{\nu }K  = k_G p_{\nu } . 
\label{c2}
\end{equation}

{Expressions} 
 (\ref{der-K2}) and (\ref{c1}) differ slightly from those found in the standard 5D Einstein
{Equations}
~(\ref{F-Einstein-5}), expressing the breaking of spacetime symmetry to
{O(3,1)}.
The differences are 
\begin{equation}
g_{\mu \nu }\left( S+\sigma\kappa\right) \rightarrow P_{\mu \nu }S , 
\end{equation}
in (\ref{der-K2}) and
\begin{equation}
-\sigma k_G \kappa \ 
\longrightarrow  \ -k_G \left( S+\sigma \kappa \right) ,  
\end{equation}
in (\ref{c1}).

In some cases, such as a diagonal metric in Cartesian coordinates, it is
possible to formulate the evolution equations directly in the quatrad frame
~\cite{Land_2023} in the simplified form
\begin{equation}
\frac{1}{c_{5}}\partial _{\tau }E_{\mu }^{\hspace{4pt}k}=-E_{\mu }^{\hspace{
4pt}l}K_{l}^{k} , 
\label{qf-1}
\end{equation}
\begin{equation}
\frac{1}{c_{5}}\partial _{\tau }K_{kl}=-\sigma \bar{R}_{kl}+KK_{kl}+\sigma 
k_G \left( S_{kl}-\frac{1}{2}\eta _{kl}S\right) , 
\label{qf-2}
\end{equation}
with constraints
\begin{equation}
\bar{R}-\sigma \left( K^{2}-K^{kl}K_{kl}\right) =-k_G \left(
S+\sigma \kappa \right) , 
\label{qf-3}
\end{equation}
\begin{equation}
D_{k}K_{m}^{k}-D_{m}K=k_G p_{m} , 
\label{qf-4}
\end{equation}
providing an initial value problem for $E_{\mu }^{\hspace{4pt}k}$ and $K_{kl}$, with~
the metric obtained from the vierbein~field.

\subsection{Weak Field~Approximation}
\label{weak}

As in standard GR, the~weak field approximation~\cite{RCM,universe8030185}
poses the local metric as a small perturbation $ h_{\alpha \beta } $ of the flat metric 
\begin{equation}
\eta _{\alpha \beta }=\text{diag}\left( -1,1,1,1,\sigma \right) , 
\label{E-24}
\end{equation}
so that
\begin{equation}
g_{\alpha \beta }=\eta _{\alpha \beta }+h_{\alpha \beta } \
\longrightarrow \ \partial _{\gamma }g_{\alpha \beta }=\partial
_{\gamma }h_{\alpha \beta }\qquad \qquad \left( h_{\alpha \beta }\right)
^{2}\simeq 0  .
\label{E-23}
\end{equation}

{In} this approximation, the~5D Ricci tensor takes the form
\begin{equation}
R_{\alpha \beta }\simeq \frac{1}{2}\left( \partial _{\beta }\partial
_{\gamma }h_{\alpha }^{\gamma }+\partial _{\alpha }\partial _{\gamma
}h_{\beta }^{\gamma }-\partial ^{\gamma }\partial _{\gamma }h_{\alpha \beta
}-\partial _{\alpha }\partial _{\beta }h\right) 
= -\frac{1}{2}\partial ^{\gamma }\partial _{\gamma }h_{\alpha \beta }  , 
\end{equation}
where $h = \eta^{\alpha\beta} h_{\alpha \beta }$ and we imposed the Lorenz
gauge condition 
\begin{equation}
\partial ^{\beta }\left( h_{\alpha \beta }-\frac{1}{2}\eta _{\alpha \beta
}h\right) =0   , 
\end{equation}
permitted by invariance of the metric under a 5D translation $x^\alpha \longrightarrow 
x^\alpha + \Lambda^\alpha (x,\tau) $.
Conveniently exploiting the Ricci tensor in this form, the~SHP field Equation~(\ref{SHP_field})
becomes the wave equation 
\begin{equation}
-\partial ^{\gamma }\partial _{\gamma } h_{\alpha \beta } = 
-\left( \partial ^{\mu }\partial _{\mu } +\sigma 
\frac{1}{c_5^2} \partial_\tau ^2 \right) h_{\alpha \beta }
= 2k_G \left( T_{\alpha \beta }-\frac{1}{2} P_{\alpha \beta }S\right)   , 
%
\label{wave}
\end{equation}
%
which admits the principal part Green's function~\cite{RCM} 
\begin{equation}
G(x,\tau ) = {\frac{1}{{2\pi }}}\delta (x^{2})\delta (\tau )+{\frac{c_{5}}{{
2\pi ^{2}}}}{\frac{\partial }{{\partial {x^{2}}}}}{\theta (-\sigma
g_{\alpha \beta }x^{\alpha }x^{\beta })}{\frac{1}{\sqrt{-\sigma
g_{\alpha \beta }x^{\alpha }x^{\beta }}}}    , 
\label{GF}
\end{equation}
%
in which the leading term, denoted $G_{\text{Maxwell}}$, has
lightlike support at equal-$\tau$ and is dominant at long distances. 
The second term, denoted $G_{\text{correlation}}$, 
drops off as $1/\text{distance}^2$ and has spacelike support
for $\sigma = -1$ or timelike support for $\sigma = +1$. 

For a general trajectory, we may consider an event distribution moving in tandem
in the neighborhood of a point with 5D coordinates  
\begin{equation}
X^{\alpha }(\tau )=(X^{\mu }\left( \tau \right) ,c_{5}\tau ) , 
\end{equation}
%
and for the shared velocity we introduce the notation 
\begin{equation}
\xi ^{\alpha }(\tau )=\frac{1}{c}u^{\alpha }\left( \tau \right) =\frac{1}{c}
\frac{dX^{\alpha }}{d\tau }  . 
\end{equation}

{The} spacetime event density is
\begin{equation}
\rho \left( x,\tau \right) =\rho \left( x-X\left( \tau \right) \right) , 
\end{equation}
leading to the mass-energy-momentum tensor 
\begin{equation}
T^{\alpha \beta }=M\rho \left( x,\tau \right) \dot{X}^{\alpha }\dot{X}
^{\beta }=M\rho \left( x,\tau \right) u^{\alpha }u^{\beta }=Mc^{2}\rho
\left( x,\tau \right) \xi ^{\alpha } (\tau) \xi ^{\beta } (\tau)  , 
\label{T-source}
\end{equation}
which is seen to be conserved by simply noting that $\;\partial _{\tau
}\rho (x,\tau )=-\xi ^{\mu }\partial _{\mu }\rho (x,\tau )$.
The generic solution for the metric perturbation is thus
\begin{equation}
h^{\alpha \beta }\left( x,\tau \right)
= 2 k_G \int
d^{4}x^{\prime }d\tau ^{\prime }G\left( x-x^{\prime },\tau -\tau ^{\prime
}\right) \left( \xi ^{\alpha } \xi ^{\beta } -\frac{1}{2}P^{\alpha \beta } \p
 \widehat \xi \p \p ^2
\right) \rho (x^\prime,\tau ^\prime) , 
%
\end{equation}
where $ \xi^\alpha = \xi^\alpha \left( \tau^\prime\right)$ and $\widehat \xi \p \p ^2 =
\widehat \eta^{\mu\nu} \xi_\mu \xi_\nu$.

\section{The Metric as Solution to a 5D Wave~Equation}
\label{5D-wave}

We are interested in the metric induced by `static' events narrowly distributed
along their $t$-axis at the spatial origin ${\mathbf x} = 0$ and evolving uniformly.
As a preliminary case, we consider a source distributed evenly along the $t$-axis in its rest
frame, as~is typically posed in standard relativity.
The center of the event trajectory is then
\begin{equation}
X(\tau) = (c\tau, {\mathbf 0}, c_5 \tau ) \ \longrightarrow \ 
\xi(\tau) = \left( 1, {\mathbf 0}, \frac{c_5  }{c} \right)  , 
\label{t-axis}
\end{equation}
and the event density 
\begin{equation}
\rho(x,\tau) =\rho \left( x-X\left( \tau \right) \right) = \rho \left( ct-c\tau
\right) \delta^{(3)}({\mathbf x}) = \delta^{(3)}({\mathbf x}) , 
\label{rho-delta}
\end{equation}
is independent of $t$ and $\tau$. 
Because $\xi(\tau) $ is constant, the~generic solution to the wave equation becomes
\begin{equation}
h_{\alpha \beta }\left( x,\tau \right) 
= 2k_G Mc^2 \ Z_{\alpha\beta}
%
\mathcal{G}\left[ \rho \left( x,\tau \right) \right] , 
\label{generic}
\end{equation}
where we denote
\begin{eqnarray}
Z_{\alpha\beta} \eq \xi_\alpha\xi_\beta - \frac{1}{2} \widehat \eta_{\alpha\beta}
\xi^\mu\xi_\mu ,  \strt{12}
\\
\mathcal{G}\left[ \rho \left( x,\tau \right) \right] \eq \int d^{4}x^{\prime
}~d\tau ^\prime G\left( x-x^{\prime },\tau -\tau ^{\prime }\right) \rho
\left( x^\prime,\tau^\prime \right)  , 
\end{eqnarray}
as kinematic and dynamic factors.
Integration of the event density (\ref{rho-delta}) with
the Green's function (\ref{GF}) leads to 
\begin{equation}
\mathcal{G}_{\text{Maxwell}}\left[ \rho \left( x,\tau \right) \right] =
\int d^4x^\prime d\tau^\prime {\frac{1}{{2\pi }}}\delta ((x-x^\prime)^{2})\delta (\tau
-\tau^\prime ) \delta^{(3)}({\mathbf x}^\prime)
= \frac{1}{4\pi \vert \mathbf{x} \vert}  , 
\end{equation}
\begin{equation}
\mathcal{G}_{\text{correlation}}\left[ \rho \left( x,\tau \right) \right] =
\frac{c_{5}}{{ 2\pi ^{2}}}
\int d^4x^\prime d\tau^\prime 
{\frac{\partial }{{\partial {x^{2}}}}}{\frac{{\theta (-\sigma
(x-x^\prime)^{\alpha }(x-x^\prime)_{\alpha })}}{\sqrt{-\sigma
(x-x^\prime)^{\alpha }(x-x^\prime)_{\alpha }}}}
\delta^{(3)}({\mathbf x}^\prime)
= 0 , 
\end{equation}
and so taking
\begin{equation}
k_G = \frac{8\pi G}{c^{4}} , 
%
%
\end{equation}
the spacetime part of the metric becomes
\begin{equation}
g_{\mu \nu }=\text{diag}\left( -1+\frac{2GM}{c^{2}r},\left( 1+\frac{2GM}{
c^{2}r}\right) \delta _{ij}\right) \simeq \text{diag}\left( -U ,U^{-1}\delta
_{ij}\right) , 
\label{sph-sym}
\end{equation}
where
\begin{equation}
U=\left( 1-\frac{2GM}{c^{2}r}\right) .
\end{equation}

{Naturally,} this metric is spatially isotropic, and~is $t$-independent because the event density is spread
evenly along the $t$-axis.  
Transforming to spherical coordinates (\ref{sph-sym}) becomes 
\begin{equation}
g_{\mu \nu }=\text{diag}\left( -U,U^{-1},U^{-1}r^{2},U^{-1}r^{2}\sin ^{2}\theta
 \right) , 
\end{equation}
which for weak fields is recognized as the Schwarzschild metric 
\begin{equation}
g_{\mu \nu }= \text{diag}\left( -U,U^{-1},R^{2},R^{2}\sin ^{2}\theta \right) , 
%
%
\end{equation}
when expressed in the isotropic coordinates~\cite{MTW} defined through
\begin{equation}
R=r\left( 1+\dfrac{k}{2r}\right)^2 . 
\end{equation}

{The} Schwarzschild metric is well-known to be Ricci-flat, $R_{\mu\nu} = 0$, a~consequence of its $R$-dependence and $t$-independence.

To study the field induced by an event localized in both space and time,
we again consider an event distribution centered on the $t$-axis around the trajectory
(\ref{t-axis}), but~write an event density
\begin{equation}
\rho \left( x,\tau \right) = \varphi \left( t-\tau \right) \delta^{(3)}\left( 
\mathbf{x}\right) \qquad \qquad \varphi_{\text{max}} = \varphi ( 0 ) , 
\end{equation}
with support in a neighborhood around $t = \tau$.
Writing $i,j=1,2,3$, the~kinematic factors are
\begin{equation}
\begin{array}{lcl}
Z_{00}= \dfrac{1}{2}
& \qquad \qquad  & 
Z_{05}= Z_{50}= -\sigma \dfrac{c_5}{c}
\rule[-18pt]{0pt}{18pt} \\ 
Z_{ij}= \dfrac{1}{2} \delta_{ij} 
& \qquad  \qquad \qquad & 
Z_{55}= \dfrac{c_5^2}{c^2} 
\end{array}  
\label{gen-sol}
\end{equation}
and the dynamic factors are
\begin{equation}
\mathcal{G}_{\text{Maxwell}}=
\int d^4x^\prime d\tau^\prime {\frac{1}{{2\pi }}}\delta ((x-x^\prime)^{2})\delta (\tau
-\tau^\prime ) \varphi \left( t^\prime-\tau^\prime \right) \delta^{(3)}({\mathbf
x}^\prime) , 
%
\end{equation}
\begin{equation}
\mathcal{G}_{\text{correlation}}=
\frac{c_{5}}{{ 2\pi ^{2}}}
\int d^4x^\prime d\tau^\prime 
{\frac{\partial }{{\partial {x^{2}}}}}{\frac{{\theta (-\sigma
(x-x^\prime)^{\alpha }(x-x^\prime)_{\alpha })}}{\sqrt{-\sigma
(x-x^\prime)^{\alpha }(x-x^\prime)_{\alpha }}}}
\varphi \left( t^\prime-\tau^\prime \right) \delta^{(3)}({\mathbf x}^\prime) \ .
\label{G-corr}
\end{equation}

{The} leading term is easily evaluated as
\begin{equation}
\mathcal{G}_{\text{Maxwell}} = \frac{\varphi \left( t-\vert \mathbf{x}\vert /c
-\tau \right) }{4\pi \vert \mathbf{x} \vert} , 
\label{leading-0}
\end{equation}
producing a gravitational field that is maximized at $\tau = t-\vert \mathbf{x}\vert/c$.
%
%
Since the source is centered at $t_{\text{source}} = \tau$, 
a test event evolving along its $t$-axis, at~a constant spatial
distance $r$ from the source, will feel the strongest gravitational
force if it is located at $t = t_{\text{source}} + \vert \mathbf{x}\vert/c$,
placing the test event on the lightcone of the source and
accounting for the propagation time of the gravitational field.
This part of the solution is comparable to the Coulomb force given in
(\ref{coulomb}) in~electrodynamics.

The evaluation for $\mathcal{G}_{\text{correlation}}$ will depend upon
the choice of $\sigma$ and the details of the distribution
$\varphi \left( s \right)$, in~nearly all cases leading to 
numerical integration.  
Taking the derivative in $\mathcal{G}_{\text{correlation}}$ we have
\begin{equation}
G_{\text{Correlation}}\left( x,\tau \right) = {\frac{c_{5}}{{2\pi ^{2}}}}
\left( \frac{1}{2}{\frac{{\theta ({-\sigma x^{2}}-c_{5}^{2}\tau ^{2})}}{\left( {
-\sigma x^{2}}-c_{5}^{2}\tau ^{2}\right) ^{3/2}}}-{\frac{{\delta }\left(
{-\sigma x^{2}}
-c_{5}^{2}\tau ^{2}\right) }{\left( {-\sigma x^{2}}-c_{5}^{2}\tau ^{2}\right) ^{1/2}
}}\right)  , 
\end{equation}
and writing
\begin{equation}
-\sigma x^2-c_5^2\tau^2 = c^2 \left[ -\sigma \left( \frac{{\mathbf x}^2}{c^2} - t^2
\right) -\frac{c_5^2}{c^2}\tau^2 \right]  , 
\end{equation}
we might consider neglecting $ c_5^2 / c^2 \ll 1$.
However, doing so makes this part of Green's function, independent of $\tau$,
so that the $\tau$ integration in (\ref{G-corr}) becomes
\begin{equation}
\int d\tau^\prime \ \varphi \left( t^\prime-\tau^\prime \right) = 1 , 
%
%
\end{equation}
and the remaining integral is
\begin{equation}
\mathcal{G}_{\text{correlation}}=
\frac{c_{5}}{{ 2\pi ^{2}}}
{\frac{\partial }{{\partial {x^{2}}}}}\int d^4x^\prime 
{\frac{{\theta (-\sigma
(x-x^\prime)^{\alpha }(x-x^\prime)_{\alpha })}}{\sqrt{-\sigma
(x-x^\prime)^{\alpha }(x-x^\prime)_{\alpha }}}}  \delta^{(3)}({\mathbf
x}^\prime)= 0 , 
%
%
\end{equation}
leaving no contribution from $G_{\text{Correlation}}$.  
In this sense, neglecting the contribution from this term is equivalent to the
$\tau$-equilibrium condition, a~point we will examine again in
Section~\ref{evolution}.

To obtain a sense of $\mathcal{G}_{\text{correlation}}$ we choose the
infinitely narrow distribution $\varphi \left( t-\tau \right) = \delta \left( t-\tau \right)$
so that
\begin{equation}
\mathcal{G}_{\text{correlation}}=
{\frac{c_{5}}{{2\pi ^{2}}}} \int ds 
\left( \frac{1}{2}{\frac{{\theta (g(s))}}{\left( g(s)\right) ^{3/2}}}-{\frac{{\delta }\left(
g(s)\right) }{\left( g(s)\right) ^{1/2}
}}\right) , 
%
%
\end{equation}
where
\begin{equation}
g(s) = c^2\left( \sigma \left( t-s
\right)^2-\sigma \dfrac{\mathbf{x}^2}{c^2}-\dfrac{c_5^2}{c^2}\left( \tau -s
\right)^2\right)  , 
%
%
\end{equation}
in which the singularities of the two integrands cancel each other out when handled carefully. 
For $\sigma = -1$, describing spacelike support, $g(s)>0$ between the
roots of $g(s) = 0$ and cancellation of singularities causes the integral to vanish.
Taking $\sigma = 1$, describing timelike support, $g(s)>0$ above the
upper root and so the integral takes its value as $s \longrightarrow \infty$
giving
\begin{equation}
\mathcal{G}_{\text{Correlation}}
= \frac{c_{5}}{2{\pi ^{2}}}\dfrac{1}{\mathbf{x}^2-{c_{5}^{2}}\tau \left( 2t-\tau
\right) } \ .
\end{equation}

{Since} this terms drops off as $1 / \mathbf{x}^2$ the contribution from $
\mathcal{G}_{\text{Maxwell}} $ will be dominant at long~distance.

In summary, the~perturbed metric is
%
\begin{eqnarray}
g_{00} \eq -U = -1+k_G M c^2\mathcal{G}\left[ \varphi \left( t-\tau \right) \delta^{(3)}\left( 
\mathbf{x}\right) \right] ,  \strt{12}
\label{g00}
\\
g_{ij} \eq V \delta_{ij} = \left( 1+k_G M c^2\mathcal{G}\left[ \varphi \left( t-\tau \right) \delta^{(3)}\left( 
\mathbf{x}\right) \right] \right) \delta_{ij} \, \qquad i,j=1,2,3\strt{12}
\label{gij}
\\
g_{05} \eq g_{50} = -2\sigma \frac{c_5}{c} k_G \mathcal{G}\left[ \varphi \left( t-\tau \right) \delta^{(3)}\left( 
\mathbf{x}\right) \right] , \strt{12}
\\
g_{55} \eq 2 \frac{c^2_5}{c^2} k_G \mathcal{G}\left[ \varphi \left( t-\tau \right) \delta^{(3)}\left( 
\mathbf{x}\right) \right] . 
\end{eqnarray}

Using (\ref{geo-eqn}) we may write the equations of motion for a nonrelativistic test event 
as
\begin{equation}
0 = \ddot{x}^{\mu }+c^{2}\left( \Gamma _{00}^{\mu }\dot{t}^{2}+2\Gamma
_{i0}^{\mu }\dfrac{\dot{x}^{i}}{c}\dot{t}+\Gamma _{ij}^{\mu }\dfrac{\dot{x}
^{i}}{c}\dfrac{\dot{x}^{j}}{c}+2\dfrac{c_{5}}{c}\Gamma _{50}^{\mu }\dot{t}+2
\dfrac{c_{5}}{c}\Gamma _{5i}^{\mu }\dfrac{\dot{x}^{i}}{c}+\dfrac{c_{5}^{2}}{
c^{2}}\Gamma _{55}^{\mu }\right) , 
%
%
\end{equation}
and from
\begin{equation}
\Gamma _{\beta \gamma }^{\mu }=\frac{1}{2}\left( \eta ^{\mu \alpha }\frac{
\partial h_{\alpha \beta }}{\partial x^{\gamma }}+\eta ^{\mu \alpha }\frac{
\partial h_{\alpha \gamma }}{\partial x^{\beta }}-\eta ^{\mu \alpha }\frac{
\partial h_{\beta \gamma }}{\partial x^{\alpha }}\right) , 
\end{equation}
evaluate the nonzero Christoffel symbols
\begin{equation}
\Gamma _{00}^{\mu }=-\dfrac{1}{2c}\delta ^{\mu 0}\frac{\partial
h_{00}}{\partial t}-\frac{1}{2}\delta ^{\mu k}\frac{\partial h_{00}}{
\partial x^{k}}
\qquad \qquad 
\Gamma _{i0}^{\mu }=\dfrac{1}{2c}\delta ^{\mu j}\frac{\partial
h_{ji}}{\partial t}-\frac{1}{2}\delta ^{\mu 0}\frac{\partial h_{00}}{
\partial x^{i}} , 
\end{equation}
\begin{equation}
\Gamma _{ij}^{\mu }=\frac{1}{2}\delta ^{\mu k}\left( \frac{\partial h_{ki}}{
\partial x^{j}}+\frac{\partial h_{kj}}{\partial x^{i}}-\frac{\partial h_{ij}
}{\partial x^{k}}\right) +\frac{1}{2c}\delta ^{\mu 0}\frac{\partial
h_{ij}}{\partial t} , 
\end{equation}
\begin{equation}
\Gamma _{50}^{\mu }=-\frac{1}{2c_{5}}\delta ^{\mu 0}\frac{
\partial h_{00}}{\partial \tau }
\qquad \qquad \qquad 
\Gamma _{5i}^{\mu }=\frac{1}{2c_{5}}\delta ^{\mu k}\frac{\partial
h_{ki}}{\partial \tau } , 
\end{equation}
where we used $h_{0i}=0, \  i=1,2,3$ and dropped $h_{5\alpha } \propto c_5 / c
\ll 1$.
Similarly neglecting terms containing $\dot{x}^{i} / c \ll 1$ 
the equations of motion split into
\begin{equation}
0=\ddot{t}-\frac{1}{2}\frac{\partial h_{00}}{\partial t}\dot{t}^{2}-\left( 
\frac{\partial h_{00}}{\partial \tau }+\mathbf{\dot{x}}\cdot \nabla
h_{00}\right) \dot{t} \qquad \qquad
0=\mathbf{\ddot{x}}-\frac{1}{2}c^{2}\dot{t}^{2}~\nabla h_{00} , 
\end{equation}
which differ from (\ref{post-N}) in the $t$-dependence of $h_{00}$.
In spherical coordinates, using 
\begin{equation}
h_{00} = h_{00} (t, r, \tau) \ \longrightarrow \ \mathbf{\dot{x}}\cdot \nabla
h_{00} = \dot r \partial_r h_{00} , 
%
%
\end{equation}
the equations of motion become
\begin{equation}
\ddot{t}=\frac{1}{2}\left( \partial_t h_{00} \right) \dot{t}^{2}+\left( 
\partial_\tau h_{00}+\dot r \partial_r h_{00}\right) \dot{t}
\qquad \qquad
\ddot{r}= \frac{1}{2}c^{2}(\partial _{r}h_{00})\dot{t}^{2}+
\frac{L^{2}}{m^2r^{3}}  , 
\label{eq-mot}
\end{equation}
where we again introduce the conserved angular momentum $L = m r^2 \dot \phi$.

{To obtain a sense of this result, we localize the source in $t$ by taking the Gaussian distribution}
\begin{equation}
\varphi \left( s\right) = \dfrac{1}{\sqrt{2\pi }\lambda _{0}} e^{-s^{2}/\lambda
_{0}^{2}} , 
%
%
\end{equation}
where $\lambda_0$ is a time scale representing the width of the event distribution
along the $t$-axis, and~consider only the leading term $\mathcal{G}_{\text{Maxwell}}$.
From (\ref{g00}) the metric takes the form
\begin{equation}
h_{00}= \dfrac{k_G Mc^2}{4\pi r} \dfrac{1}{\sqrt{2\pi }\lambda _{0}}
\exp \left[ -\dfrac{\left( t- r / c -\tau \right) ^{2}}{\lambda
_{0}^{2}}\right]
= \dfrac{k_G Mc^2}{4\pi} \dfrac{1}{\sqrt{2\pi }\lambda _{0}}
\frac{1}{r}\hat \varphi , 
%
%
\end{equation}
where for convenience we notate
\begin{equation}
\hat \varphi = \exp \left[ -\dfrac{\left( t- r / c -\tau \right) ^{2}}{\lambda
_{0}^{2}}\right] . 
%
%
\end{equation}

{The} partial derivatives are
\begin{equation}
\partial_t \p h_{00}= - \dfrac{k_G Mc^2}{4\pi} \dfrac{1}{\sqrt{2\pi }\lambda _0 }
\frac{2\left( t- r / c -\tau \right)}{\lambda_0^2}
\frac{1}{r}\hat \varphi , 
%
%
\end{equation}
\begin{equation}
\partial_\tau \p h_{00}=  \dfrac{k_G Mc^2}{4\pi} \dfrac{1}{\sqrt{2\pi }\lambda _0 }
\frac{2\left( t- r / c -\tau \right)}{\lambda_0^2}
\frac{1}{r}\hat \varphi , 
%
%
\end{equation}
\begin{equation}
\partial_r \p h_{00}=  \dfrac{k_G Mc^2}{4\pi} \dfrac{1}{\sqrt{2\pi }\lambda _0 }
\left[ 
-\frac{1}{r} + 
\frac{2\left( t- r / c -\tau \right)}{\lambda_0^2 }
\right] 
\frac{1}{r}\hat \varphi , 
%
%
\end{equation}
leading to the equations of motion
\begin{equation}
\ddot{t}=\dfrac{k_{G}Mc^{2}}{4\pi }\dfrac{1}{\sqrt{2\pi }\lambda _{0}}\left[
\frac{2\left( t-r/c-\tau \right) }{\lambda _{0}^{2}}\left( -\frac{1}{2}\dot{t
}^{2}+\dot{t}+\dfrac{\dot{r}}{c}\right) \frac{1}{r}-\frac{\dot{r
}}{r^{2}}\right] \hat{\varphi} , 
\end{equation}
\begin{equation}
\ddot{r} = \frac{1}{2}\left( \dfrac{k_{G}Mc^{2}}{4\pi }\dfrac{c}{\sqrt{2\pi }
\lambda _{0}}\left[ -\frac{1}{r}+\frac{2\left( t-r/c-\tau \right) }{\lambda
_{0}^{2}}\right] \frac{1}{r}\hat{\varphi}\right) \dot{t}^{2}+\frac{L^{2}}{
m^{2}r^{3}} \ . 
\end{equation}

{Locating} the test event on the lightcone of the source event
\begin{equation}
t-\frac{r}{c} -\tau = 0 \ \longrightarrow \ \hat{\varphi}=1 , 
%
%
\end{equation}
the equations of motion reduce to
\begin{equation}
\ddot{t}=-\dfrac{k_{G}Mc^{2}}{4\pi r^{2}}\dfrac{1}{\sqrt{2\pi }\lambda _{0}}
\ \frac{\dot{r}}{c} , 
\label{t-eqn}
\end{equation}
\begin{equation}
\ddot{r}=-\frac{1}{2}\dfrac{k_{G}Mc^{2}}{4\pi }\dfrac{c}{\sqrt{2\pi }
\lambda _{0}}\frac{1}{r^{2}}\dot{t}^{2}+\frac{L^{2}}{m^{2}r^{3}} \ .
\label{r-eqn}
\end{equation}

{Since} we must have $\dot r / c \longrightarrow 0 $ in (\ref{t-eqn}) we may write $\dot t = 1$
which recovers Newtonian gravitation in (\ref{r-eqn}) by putting
\begin{equation}
\frac{1}{2}\dfrac{k_{G}Mc^{2}}{4\pi }\dfrac{c}{\sqrt{2\pi }\lambda _{0}}=GM
\qquad  \longrightarrow \qquad  k_{G}=\sqrt{2\pi } \p \dfrac{8\pi G}{c^{2}}\p
\dfrac{\lambda _{0}}{c} , 
\end{equation}
in which the inverse length $\lambda_0 / c$ 
compensates for the dimensions $1/\text{length}^4$ of the spacetime event
density, in~relation to the usual $1/\text{length}^3$ dimensions of 
particle density.
For an arbitrary position of the test event on the $t$-axis with  $\dot r / c \ll 1$, the~metric
perturbation and equations of motion are 
\begin{equation}
h_{00}=  \dfrac{2 GM}{c^{2}r} \hat{\varphi} \ \longrightarrow \ 
g_{00}=  -\left( 1 - \dfrac{2 GM}{c^{2}r} \hat{\varphi} \right)  , 
%
%
\end{equation}
\begin{equation}
\ddot{t}=\dfrac{2 GM}{c^{2}}\left[
\frac{2\left( t-r/c-\tau \right) }{\lambda _{0}^{2}}\left( -\frac{1}{2}\dot{t
}^{2}+\dot{t}
\right) \frac{1}{r}-\frac{\dot{r
}}{r^{2}}\right] \hat{\varphi} , 
\label{t-f}
\end{equation}
\begin{equation}
\ddot{r} = - \frac{GM}{r^2} \left[ 1 -\frac{r}{\lambda_0 c}\frac{2\left( t-r/c-\tau \right) }{\lambda
_0}\right] \dot{t}^{2} \ \hat{\varphi}+\frac{L^{2}}{
m^{2}r^{3}} \ . 
\label{rad-f}
\end{equation}

{For} the nonrelativistic event, 
the $t$ equation can be approximated
\begin{equation}
\ddot{t} \simeq \dfrac{2 GM}{c^{2}r}\left[
\frac{ t-r/c-\tau  }{\lambda _{0}^{2}}-\frac{\dot{r
}}{r}\right] \hat{\varphi} , 
\end{equation}
which is a product of small factors, so that acceleration in time will remain
negligible.
But the radial equation depends on the ratio $r / \lambda_0 c$ of the radial distance, taken
to be large, and~the width of the $t$ distribution.
Equation~(\ref{rad-f}) approximates Newtonian gravitation for $t-r/c-\tau = 0$,
but as the test event accelerates in the radial direction under the resulting
force, the~distance will decrease as $r \longrightarrow r - \delta r$ and so the
acceleration becomes
\begin{equation}
\ddot{r} \simeq - \frac{GM}{r^2} \left[ 1 -2\frac{r}{\lambda_0 c}\frac{ \delta r }{\lambda
_0 c}\right]  \hat{\varphi}+\frac{L^{2}}{
m^{2}r^{3}} \ . 
%
%
%
\end{equation}

{This} shows that the width $\lambda_0$ of the source event distribution along the
$t$-axis must be much larger than $r / c$, or~else the gravitational force will
weaken and possibly change sign from attraction to repulsion.  
In the limit $\lambda_0 \longrightarrow \infty$, we have $\hat{\varphi} = 1$ and
the metric (\ref{g00}) and (\ref{gij}) recover the $t$-independent metric
(\ref{sph-sym}), losing the $t$-localization.  

The resulting model, which is less than adequate, follows from a series of
approximations, in~particular using the linearized 5D theory and neglecting
$\mathcal{G}_{\text{correlation}}$. 
Although we used a particular distribution $\varphi(t,r,\tau)$ to arrive at the
equations of motion,  any solution to the 5D wave equation found from the
Green's function 
will have the form 
\begin{equation}
h_{00} \propto \frac{1}{r} \hat \varphi(t,r,\tau) , 
\end{equation}
as its leading term.
As a result, the~gravitational force appearing in the radial Equation~(\ref{eq-mot})
will take the form 
\begin{equation}
\partial_r h_{00} \propto -\frac{1}{r^2}\big[  \hat \varphi(t,r,\tau)
- r \partial_r \hat \varphi(t,r,\tau)\big]  , 
\label{sufficient}
\end{equation}
which may change the sign for any narrow distribution with $\partial_r \hat \varphi$
sufficient large at some value of its argument.
For example, using the distribution
\begin{equation}
 \hat \varphi(x,\tau) = \frac{1}{2}e^{-\left\vert t - \vert {\mathbf x} \vert
/ c - \tau \right\vert / \lambda_0}
\qquad \longrightarrow \qquad r\partial_r  \hat \varphi = \frac{r}{\lambda_0}
\func{sgn}\left( t - \vert {\mathbf x} \vert / c - \tau \right)  , 
%
%
\end{equation}
the gravitational force may change sign sharply for a small shift in the
$\tau$-synchronization of the test particle around $\tau = t - \vert {\mathbf x}
\vert / c$.
It thus appears that linearized GR will not provide an adequate model for the
localized metric produced by a localized event.
In Section~\ref{evolution}, we analyze this question further in the context of
the 4+1 method, and~show that the initial value problem in full nonlinear GR
involves a more complex structure than revealed in the 5D wave equation
approach.

\section{The Metric as Solution to 4+1 Evolution~Equations}
\label{evolution}

In this section, we apply the {4+1} method to study the solution to the linearized
field equations found in Section~\ref{5D-wave}. 
As mentioned in Section~\ref{local}, the~initial value problem may be posed in
the quatrad frame using the simplified Equations~(\ref{qf-1})--(\ref{qf-4}) because the
spacetime part of the metric is diagonal.
It is easily seen that for 
\begin{equation}
\gamma _{\mu \nu }=\func{diag}\left( -U,V,V,V\right) , 
\label{v-1}
\end{equation}
%
the vierbein field takes the form
\begin{equation}
E_{\mu }^{\hspace{4pt}k}=\sqrt{U}\delta _{\mu }^{\hspace{4pt}0}\delta _{0}^{
\hspace{4pt}k}+\sqrt{V}\delta _{\mu }^{\hspace{4pt}s}\delta _{s}^{\hspace{4pt
}k} \qquad \qquad 
e_{\hspace{4pt}k}^{\mu }=\dfrac{1}{\sqrt{U}}\delta _{\hspace{4pt}0}^{\mu
}\delta _{\hspace{4pt}k}^{0}+\dfrac{1}{\sqrt{V}}\delta _{\hspace{4pt}s}^{\mu
}\delta _{\hspace{4pt}k}^{s} , 
\label{v-2}
\end{equation}
%
where $s,t=1,2,3$.
For the event distribution along the $t$-axis, the~mass-energy-momentum tensor
\begin{equation}
T^{\alpha \beta }=Mc^{2}\varphi \left( t-\tau \right) \delta ^{\left(3\right)
}\left( \mathbf{x}\right) \left( \delta _{0}^{\alpha }+ \frac{c_5}{c}\delta
_{5}^{\alpha }\right) \left( \delta _{0}^{\beta }+ \frac{c_5}{c}\delta_{5}^{\beta
}\right)  , 
\end{equation}
decomposes to
\begin{eqnarray}
\label{kappa}
\kappa  \eq T^{55}=\frac{c_5^2}{c^2} \p mc^2{\varphi}\left( t-\tau \right) \delta
^{\left( 3\right) }\left( \mathbf{x}\right)  , 
\\
p_{k}  \eq -\sigma \eta _{kk^{\prime }}E_{\mu }^{\hspace{4pt}k^{\prime
}}T^{5\mu } , 
=\frac{c_5}{c}\sigma \sqrt{U} \p  mc^{2}{
\varphi}\left( t-\tau \right) \delta ^{\left( 3\right) }\left( \mathbf{x}
\right) ,  
\\
S_{kl}  \eq \eta _{kk^{\prime }}\eta _{ll^{\prime }}E_{\mu
}^{\hspace{4pt}k^{\prime }}E_{\mu ^{\prime }}^{\hspace{4pt}l^{\prime }}T^{\mu
\mu ^{\prime}}=
\eta _{k0}\eta _{l0}
U 
 \p mc^{2}{\varphi}\left( t-\tau \right)\delta ^{\left( 3\right) }\left(
\mathbf{x}\right)   , 
\\
S  \eq \eta^{kl} S_{kl}
= - Umc^{2}{\varphi}\left( t-\tau \right)
\delta^{\left( 3\right) }\left( \mathbf{x}\right) , 
\end{eqnarray}
so that 
the source for (\ref{qf-2})
\begin{equation}
S_{kl}-\frac{1}{2}\eta _{kl}S
= \left( \eta _{k0}\eta _{l0}
+\frac{1}{2}\eta _{kl} \right) Umc^{2}{\varphi}\left( t-\tau \right)
\delta^{\left( 3\right) }\left( \mathbf{x}\right)
= \frac{1}{2}\delta_{kl} S_{00} , 
%
\label{v-src}
\end{equation}
is diagonal and identical in each~component.

{In the weak field approximation, assuming a metric of the type
obtained by perturbation}
\begin{equation}
U=1-\Phi \qquad \qquad V=1+\Phi  , 
\end{equation}
entails
\begin{equation}
\sqrt{U}=\sqrt{1-\Phi }\simeq 1-\dfrac{1}{2}\Phi \qquad \qquad \sqrt{V}=\sqrt{
1+\Phi }\simeq 1+\dfrac{1}{2}\Phi \ .
\end{equation}

{Now} the vierbein field can be written
\begin{equation}
E_{\mu }^{\hspace{4pt}k} = \left( 1-\dfrac{1}{2}\Phi \right) \delta _{\mu
}^{\hspace{4pt}0}\delta _{0}^{\hspace{4pt}k}+\left( 1+\dfrac{1}{2}\Phi
\right) \delta _{\mu }^{\hspace{4pt}s}\delta _{s}^{\hspace{4pt}k}=
\delta _{\mu }^{\hspace{4pt}k}+\dfrac{1}{2} \left( -\delta _{\mu }^{
\hspace{4pt}0}\delta _{0}^{\hspace{4pt}k}+\delta _{\mu }^{\hspace{4pt}
s}\delta _{s}^{\hspace{4pt}k}\right)\Phi  , 
\end{equation}
with derivatives
\begin{equation}
\partial_\alpha E_{\mu }^{\hspace{4pt}k}=\dfrac{1}{2}\left( -\delta
_{\mu }^{\hspace{4pt}0}\delta _{0}^{\hspace{4pt}k}+\delta _{\mu }^{\hspace{
4pt}s}\delta _{s}^{\hspace{4pt}k}\right)\partial_\alpha \Phi \ . 
\end{equation}

{Since} the extrinsic curvature $K_{kl}$ must also arise as a perturbation, we
discard terms of the type $\Phi \p K_{kl} \simeq 0$ and the
first evolution Equation~(\ref{qf-1}) reduces to 
\begin{equation}
\frac{1}{2c_{5}} \left( -\delta _{\mu }^{\hspace{4pt}
0}\delta _{0}^{\hspace{4pt}k}+\delta _{\mu }^{\hspace{4pt}s}\delta _{s}^{
\hspace{4pt}k}\right)\partial _{\tau }\Phi = -\delta _{\mu
}^{\hspace{4pt}l}K_{l}^{k} , 
\end{equation}
so that lowering the $k$ index provides
\begin{equation}
%
\frac{1}{2c_{5}} \delta _{k l} \p \partial _{\tau }\Phi
= -K_{kl} \ . 
\label{e-1}
\end{equation}

{Similarly} discarding the term $K \p K_{kl} \simeq 0$, the~second evolution
Equation~(\ref{qf-2}) reduces to 
\begin{equation}
\frac{1}{c_{5}}\partial _{\tau } K_{kl} = -\sigma \bar{R}_{kl}+\frac{1}{2}\sigma
k_G  \delta _{kl} S_{00}  , 
%
\label{e-2}
\end{equation}
where we used (\ref{v-src}) as the source.
These expressions provide a pair of coupled first-order equations for 
$\Phi(x,\tau)$ and $ K_{kl}(x,\tau)$, given initial conditions 
$\Phi(x,0)$ and $ K_{kl}(x,0)$.  
It was shown in~\cite{universe8030185} that for weak fields,
the constraints (\ref{qf-3}) and (\ref{qf-4}) are equivalent to the wave
equation for $h_{5\alpha}$ and so, given the product structure of
(\ref{generic}),
these will be satisfied for any solution to
the 5D wave equation for $h_{00}$.

Using the convenient linearized form 
\begin{equation}
\bar{R}_{kl}=-\frac{1}{2} \delta _{kl} \p \partial ^{\mu }\partial _{\mu } h_{00}
=-\frac{1}{2} \delta _{kl} \p \partial ^{\mu }\partial _{\mu }\Phi , 
\label{Ric-lin}
\end{equation}
to evaluate the Ricci tensor, we see that each term in the evolution equations
is diagonal,
reducing the system to an initial value problem for $\Phi(x,\tau)$.  
By solving (\ref{e-1}) for
\begin{equation}
K_{kl} = -\frac{1}{2c_{5}} \delta _{kl} \p \partial _{\tau }\Phi , 
\end{equation}
and inserting this into (\ref{e-2}) 
we obtain
\begin{equation}
\partial ^{\mu }\partial _{\mu }\Phi + 
\sigma \frac{1}{c_{5}^{2}}\partial _{\tau }^{2}\Phi 
+ k_G S_{00}  = 0 , 
\end{equation}
which simply recovers the 5D wave Equation~(\ref{wave}) we 
analyzed in Section~\ref{5D-wave}.
Since no complete closed-form solution is readily available, we studied the leading
term 
\begin{equation}
\Phi (x,\tau) = \mathcal{G}_{\text{Maxwell}} = \frac{\varphi \left( t-\vert
\mathbf{x}\vert /c -\tau \right) }{4\pi \vert \mathbf{x} \vert}  , 
\label{leading}
\end{equation}
%
as a partial solution, and~found that the resulting geodesic equations for a
test event placed an unreasonable condition on the event density $\varphi$. 
We also showed that neglecting the subdominant terms $\mathcal{G}_{\text{correlation}}$
is equivalent to taking the limit $c_5/c \longrightarrow 0$. 
To see this another way, we rewrite the evolution Equation~(\ref{e-2}) as
\begin{equation}
\partial _{\tau } K_{kl} = \frac{1}{2} \sigma \p \delta_{kl} \w c_{5} \left[
\partial ^{\mu }\partial _{\mu }\Phi 
+ mc^{2}{\varphi}\left( t-\tau \right)
\delta^{\left( 3\right) }\left( \mathbf{x}\right) \right]  \ .
\end{equation}
leading to the equilibrium condition $\partial _{\tau } K_{kl} = 0$ either in the limit $c_5 \longrightarrow 0$,
or by setting $\Phi = \mathcal{G}_{\text{Maxwell}}$ for which
the expression in parentheses gives zero by the 4D wave equation.
As seen from (\ref{sufficient}), these problems will be present in any solution to
the linearized field equations for the source (\ref{v-src}). 

While the leading term $G_{\text{Maxwell}}$ of the Green's function 
provides adequate solutions in SHP electromagnetism, this appears
not to be the case in GR.
It appears that the model of localized events interacting through a localized
metric must be posed in the full nonlinear field theory, which admits
structures not captured by the linearized equations.
%
That is, we write the exact evolution equations and constraints (\ref{qf-1})--(\ref{qf-4}) to find a diagonal metric
(\ref{v-1}) derived from the vierbein field (\ref{v-2}).
But in the absence of linearization, the~convenient expression (\ref{Ric-lin})
for $R_{\mu\nu}$ is no longer applicable, adding significant complexity to the
problem.

As mentioned in Section~\ref{5D-wave}, the~Ricci flatness $R_{\mu\nu} = 0$ of
the Schwarzschild solution depends on the metric being a function of the
three spatial coordinates (through \hbox{$R =\vert \mathbf{x} \vert$}) but
\hbox{$t$-independent}.  
However, the~Ricci tensor for a general diagonal metric with
functional dependence on all spacetime coordinates $x^\mu$ will
necessarily have nonzero off-diagonal components~\cite{win1996ricci}.  
Thus, we may specify the initial vierbein field $E_{\mu }^{\hspace{4pt}k}(x,0)$ and
extrinsic curvature $K_{kl}(x,0)$ to be diagonal, as~is the source
(\ref{v-src}), but~$K_{kl}(x,\tau)$ will acquire off-diagonal terms from $\bar R_{kl}$ 
under the evolution described by Equation~(\ref{qf-2}).
Therefore, the~vierbein field $E_{\mu }^{\hspace{4pt}k}(x,\tau)$ will acquire
off-diagonal terms from the extrinsic curvature via Equation~(\ref{qf-1}).
As a result, the~metric $\gamma_{\mu\nu} = \eta_{kl} E_{\mu
}^{\hspace{4pt}k}E_{\nu }^{\hspace{4pt}l}$ may acquire off-diagonal terms,
in which case, \mbox{(\ref{qf-1}) and (\ref{qf-2})} will no longer be valid,
forcing us to use the coordinate frame expressions (\ref{der-gamma}) and
(\ref{der-K2}) as the evolution equations for the~metric. 

In summary, we require a metric that reproduces Newtonian gravitation for a nonrelativistic
test event at large distance, falls off to $\eta_{\mu\nu}$ as $1 / r $,
is localized around $\tau = t - \vert \mathbf{x} \vert / c$, but~is
not of the separable form (\ref{leading}). 
In addition, the~initial conditions for $\gamma_{\mu\nu}$ and $K_{\mu\nu}$ must be
chosen carefully to satisfy the constraints (\ref{c1}) and (\ref{c2}).
This list of requirements is complex and perhaps cannot be satisfied.
A subsequent paper will discuss these issues at greater~length.

\section{Conclusions and~Discussion}
\label{conclusion}

After reviewing the basic structure of the Stueckelberg--Horwitz--Piron formalism
in relativity and its extension to GR, we constructed a model in which a
localized spacetime event evolving with the invariant parameter $\tau$ 
induces a metric that similarly evolves with $\tau$. 
Extending developments in SHP electrodynamics, we fixed the event at the spatial
origin of its rest frame, in~a narrow distribution moving uniformly along its
$t$-axis, and~using the Green's function $G(x,\tau)$, solved the 5D wave
equation describing weak gravitation in linearized GR.  
The resulting solution, dominated by the leading term in $G(x,\tau)$, was shown
to be analogous to the electromagnetic Coulomb force, falling off to the flat
metric as $1/r$ and localized around the retarded time $\tau = t - r
/c$ for a test event with coordinates $x = (ct,{\mathbf x})$. 
In this picture, a~localized event produces a localized field that acts on a
remote localized event, with~the interaction synchronized by $\tau$.
However, unlike the electrodynamic Lorentz force, the~effect of the metric
through the geodesic equations of motion leads to a possible reversal of the
gravitational force, because~the functional dependence of the metric is a
separable product of $1/r$ and the localized distribution $\hat \varphi (t - r
/c -\tau)$.
This issue was shown to be a necessary feature of any solution for weak
gravitation, produced by the leading term in $G(x,\tau)$ for any source
distribution.

We conclude that while the leading term $G_{\text{Maxwell}}$ of the Green's
function provides adequate solutions in SHP electromagnetism, additional work
will be required to extend the model of a source localized in spacetime to GR.
The required metric must reproduce Newtonian gravitation for a nonrelativistic
test event at large distance, fall off to $\eta_{\mu\nu}$ as $r \rightarrow
\infty$, be localized around $\tau = t_{\text{retarded}}$, but~not be
separable.
Such a metric will likely include off-diagonal components and must be approached
in the 4+1 formalism, which poses an initial value problem for the metric.
This list of requirements is complex and whether they can be satisfied remains an open question.
Candidates for such a metric will be discussed in a subsequent~paper.







%


\begin{thebibliography}{999}

\bibitem{AS}
Land, M.
\newblock Local metric with parameterized evolution.
\newblock {\em Astron. Nachrichten} {\bf 2019}, {\em 340},~983--988.
\newblock {\url{https://doi.org/10.1002/asna.201913719}}.

\bibitem{sym12101721}
Land, M.
\newblock A 4+1 Formalism for the Evolving Stueckelberg-Horwitz-Piron Metric.
\newblock {\em Symmetry} {\bf 2020}, {\em 12}, 1721.
\newblock {\url{https://doi.org/10.3390/sym12101721}}.

\bibitem{Land_2021}
Land, M.
\newblock A new approach to the evolving 4$+$1 spacetime metric.
\newblock {\em J. Phys. Conf. Ser.} {\bf 2021}, {\em
  1956},~012010.
\newblock {\url{https://doi.org/10.1088/1742-6596/1956/1/012010}}.

\bibitem{universe8030185}
Land, M.
\newblock Weak Gravitation in the 4+1 Formalism.
\newblock {\em Universe} {\bf 2022}, {\em 8}, 185.
\newblock {\url{https://doi.org/10.3390/universe8030185}}.

\bibitem{Land_2023}
Land, M.
\newblock A vielbein formalism for SHP general relativity.
\newblock {\em J. Phys. Conf. Ser.} {\bf 2023}, {\em
  2482},~012006.
\newblock {\url{https://doi.org/10.1088/1742-6596/2482/1/012006}}.

\bibitem{Fock}
Fock, V.
\newblock Proper time in classical and quantum mechanics.
\newblock {\em Phys. Z. Sowjetunion} {\bf 1937}, {\em 12},~404--425.

\bibitem{Stueckelberg-1}
Stueckelberg, E.
\newblock La signification du temps propre en m{\'e}canique: Ondulatoire.
\newblock {\em Helv. Phys. Acta} {\bf 1941}, {\em 14},~321--322.
\newblock (In French)

\bibitem{Stueckelberg-2}
Stueckelberg, E.
\newblock Remarque a propos de la cr{\'e}ation de paires de particules en
  th{\'e}orie de relativit{\'e}.
\newblock {\em Helv. Phys. Acta} {\bf 1941}, {\em 14},~588--594.
\newblock (In French)

\bibitem{HP}
Horwitz, L.; Piron, C.
\newblock Relativistic Dynamics.
\newblock {\em Helv. Phys. Acta} {\bf 1973}, {\em 48},~316--326.

\bibitem{bound-1}
Horwitz, L.; Lavie, Y.
\newblock Scattering theory in relativistic quantum mechanics.
\newblock {\em Phys. Rev. D} {\bf 1982}, {\em 26},~819--838.
\newblock {\url{https://doi.org/doi:10.1103/PhysRevD.26.819.}}

\bibitem{bound-2}
Arshansky, R.; Horwitz, L.
\newblock Relativistic potential scattering and phase shift analysis.
\newblock {\em J. Math. Phys.} {\bf 1989}, {\em 30},~213.
\newblock {\url{https://doi.org/doi:10.1063/1.528572.}}

\bibitem{bound-3}
Arshansky, R.; Horwitz, L.
\newblock Covariant phase shift analysis for relativistic potential scattering.
\newblock {\em Phys. Lett. A} {\bf 1988}, {\em 131},~222--226.

\bibitem{bound-4}
Arshansky, R.; Horwitz, L.
\newblock The quantum relativistic two-body bound state. {I}. The spectrum.
\newblock {\em J. Math. Phys.} {\bf 1989}, {\em 30},~66.
\newblock {\url{https://doi.org/doi:10.1063/1.528591.}}

\bibitem{bound-5}
Arshansky, R.; Horwitz, L.
\newblock The quantum relativistic two-body bound state. {II}. The induced
  representation of SL (2, C).
\newblock {\em J. Math. Phys.} {\bf 1989}, {\em 30},~380.
\newblock {\url{https://doi.org/doi:10.1063/1.528456.}}

\bibitem{saad}
Saad, D.; Horwitz, L.; Arshansky, R.
\newblock Off-shell electromagnetism in manifestly covariant relativistic
  quantum mechanics.
\newblock {\em Found. Phys.} {\bf 1989}, {\em 19},~1125–1149.

\bibitem{rel-qm}
Horwitz, L.P.
\newblock {\em Relativistic Quantum Mechanics}; {Springer: Dordrecht,
  Netherlands,  2015.}
\newblock {\url{https://doi.org/10.1007/978-94-017-7261-7}}.

\bibitem{manybody}
Horwitz, L.P.; Arshansky, R.I.
\newblock {\em Relativistic Many-Body Theory and Statistical Mechanics};
  Morgan \& Claypool Publishers: {San Rafael, CA, USA, } 
2018; pp. {2053--2571.} 
\newblock {\url{https://doi.org/10.1088/978-1-6817-4948-8}}.

\bibitem{RCM}
Land, M.; Horwitz, L.P.
\newblock {\em Relativistic Classical Mechanics and Electrodynamics}; Morgan
  \& Claypool Publishers: {San Rafael, CA,} 
 \linebreak  USA, 2020.

\bibitem{SHPGR}
{Horwitz, L.P.}
\newblock {An Elementary Canonical Classical and Quantum Dynamics for General
  Relativity.}
\newblock {\em {J. Phys. Conf. Ser.}} {\bf {2019}}, {\em
  {1239}},~{012014.}
\newblock {\url{https://doi.org/10.1088/1742-6596/1239/1/012014}}.

\bibitem{SHPGR2}
{Horwitz, L.P.}
\newblock {An elementary canonical classical and quantum dynamics for general
  relativity.}
\newblock {\em  {Eur. Phys. J. Plus}} {\bf {2019}}, {\em {134}},~{313.}
\newblock {\url{https://doi.org/10.1140/epjp/i2019-12689-7}}.

\bibitem{wheeler_bio}
Wheeler, J.A.
\newblock {\em Geons, Black Holes and Quantum Foam: A Life in Physics}; W. W.
  Norton \& Company: {New York, NY, USA,}  2000.

\bibitem{Yepez}
Yepez, J.
\newblock {Einstein's vierbein field theory of curved space.} \emph{arXiv} {\bf 2011}, arXiv:1106.2037.

\bibitem{Gourgoulhon}
Gourgoulhon, E.
\newblock \emph{3+1 Formalism and Bases of Numerical Relativity};
\newblock Technical report; Laboratoire Univers et Theories, C.N.R.S.: {Paris, France,} 
  2007;
\newblock {Lectures given at the General Relativity Trimester held in the
  Institut Henri Poincare (Paris, Sept.-Dec. 2006) and at the VII Mexican
  School on Gravitation and Mathematical Physics (Playa del Carmen, Mexico, 26
  November--2 December 2006).} 


\bibitem{Bertschinger}
Bertschinger, E.
\newblock \emph{Hamiltonian Formulation of General Relativity};
\newblock Technical Report Physics 8.962; Massachusetts Institute of
  Technology: {Cambridge, MA, USA,} 2002.

\bibitem{Blau}
Blau, M.
\newblock \emph{Lecture Notes on General Relativity};
\newblock Technical Report; Albert Einstein Center for Fundamental Physics,
  Universität Bern: {Bern, Switzerland,}  2020.

\bibitem{ADM}
Arnowitt, R.L.; Deser, S.; Misner, C.W.
\newblock Republication of: The dynamics of general relativity.
\newblock {\em Gen. Relativ. Gravit.} {\bf 2004}, {\em
  40},~1997--2027.

\bibitem{Weinberg}
Weinberg, S.
\newblock {\em {Gravitation and Cosmology: Principles and Applications of the
  General Theory of Relativity}}; Wiley: New York, NY,  1972.

\bibitem{MTW}
{Misner}, C.W.; {Thorne}, K.S.; {Wheeler}, J.A.
\newblock {\em {Gravitation}}; W.H.~Freeman and Co.:  San Francisco, CA, USA,  1973.

\bibitem{win1996ricci}
Win, K.Z.
\newblock Ricci Tensor of Diagonal Metric. \emph{arXiv}  \textbf{1996}, arXiv:gr-qc/9602015


\end{thebibliography}
\end{document}